# Thermalization of one-dimensional photon gas and thermal lasers in erbium-doped fibers

**Rafi Weill, Alexander Bekker, Boris Levit, Michael Zhurahov and Baruch Fischer**

The Andrew & Erna Viterbi Faculty of Electrical Engineering, Technion, Haifa 32000, Israel

E-mail: fischer@ee.technion.ac.il

**Abstract:**

We demonstrate thermalization and Bose-Einstein (BE) distribution of photons in standard erbium–doped fibers (edf) in a broad spectral range up to ~200*nm* at the 1550*nm* wavelength regime. Our measurements were done at a room temperature ~300*K* and 77*K*. It is a special demonstration of thermalization of photons in fiber cavities and even in open fibers. They are one-dimensional (1D), meters-long, with low finesse, high loss and small capture fraction of the spontaneous emission. Moreover, we find in the edf cavities coexistence of thermal equilibrium (TE) and lasing without an <u>overall</u> inversion. The experimental results are supported by a theoretical analysis based on the rate equations.

# 1. INTRODUCTION

It is commonly accepted that photons in lasers are not in thermal equilibrium (TE), and don't show Bose-Einstein spectral distribution and Bose-Einstein condensation (BEC). We show here thermalization of photons in standard one-dimensional (1D) erbium–doped fiber (edf) cavities, including in the lasing regime, and in open fibers. Photon thermalization and photon-BEC were demonstrated in a dye-filled microcavity [1,2]. They were observed not in the lasing regime but much below it, and required strict conditions [1,2] that included a micron-size cavity with a two-dimensional (2D) confinement of lateral modes, very high mirror reflectivities that provided high finesse to trap the photons in a "white wall photon box", and very low losses. They also needed a very high capture of the spontaneously emitted photons in all directions and a cutoff frequency. All these requirements were crucial for TE and BEC. Insightful theoretical studies discussed the main differences between a low loss quantum-statistics regime that can yield TE and BEC, and a higher losses regime of classical lasers [3,4]. There were also discussions and questions about the nature of photon-BEC in optical cavities, the relation to lasers [3-8] and to classical condensation [9-17]. In our work we find that for edf systems many restrictions that were required in the microcavity experiment [1,2] are relaxed, and photon TE is obtained in standard one-dimensional (1D) edf cavities, in the lasing regime, and even in open fibers [5]. We note that the lasing occurs where the overall population of the second level is lower than the first one and therefore it can be regarded as thermal lasing without an overall inversion (T-LWI), as can be the situation in the microcavity experiment where the population buildup was regarded as BEC [1,2]. We nevertheless emphasize, that there is inversion at the specific lasing line that is pushed to a high wavelength regime in the edf spectrum due to thermalization where the thermal dependent emission rate is larger than the absorption one. We also stress that thermalization and the needed density of light-mode states (DOS) that we discuss in this paper are the important conditions for obtaining BEC, but we do not give here yet a final conclusion on its observation.

# 2. THE ERBIUM-DOPED FIBER (EDF) PLATFORM

Erbium-doped fibers (edf) are widely used as amplifiers in fiber optic communication. Erbium is a "three level" atom system with broad levels due to Stark splittings (Stark manifolds) that can provide gain at the 1550nm wavelength regime, commonly between 1530-1560*nm* (the C-band) [18]. The pumping from the first to the third level is usually done with wavelengths at ~980*nm* or

~1470 *nm*. The spontaneous emission time between the two main levels is $t_{sp} \sim 10 ms$. The absorption and emission wavelength dependent cross sections $\sigma_{12}(\lambda)$ and $\sigma_{21}(\lambda)$, which are related to transitions between the first and second levels, are shown in Fig. 1. These cross sections usually follow the McCumber [18-20] (or Kennard-Stepanov [21,22]) relation [16] $\sigma_{12}(\nu)/\sigma_{21}(\nu) = B_{12}(\nu)/B_{21}(\nu) \propto e^{\beta h\nu}$, where $\beta = 1/k_B T$, T is temperature, $h$ and $k_B$ are the Planck and Boltzmann constants, and $B_{ij}(\nu)$ are the Einstein coefficients (with their possible extended definition in several former works [16-17] that follows the temperature dependent $\sigma_{ij}(\nu)$), $\nu$ and $\lambda$ are frequency and wavelength, $\nu\lambda = c/n_r$, $c$ is the speed of light, and $n_r$ is the refractive index. We emphasize that the usual Einstein relation $B_{12} = B_{21}$ (and more generally, $g_1 B_{12} = g_2 B_{21}$, where $g_i$ is the atomic levels degeneracy factor) holds for a specific wavelength transition. The temperature dependent asymmetry in $\sigma_{12}(\nu) \neq \sigma_{21}(\nu)$ results from the sublevels thermal occupations. The broad sublevels of each of these two levels thermalize in a time scale of $t_a \sim 0.2 - 0.7 ps$ due to interaction with the silica host [18]. Without photon-photon interaction the photon thermalization depends on interaction with the erbium atoms. The time scale for such successive photon-erbium scattering is $t_p$ and the photon lifetime in the cavity is $t_c$. In our fiber systems $t_p \ll t_c$, as we discuss below, and therefore there are many photon-erbium interaction cycles that are needed for photon thermalization.

The photon population in pumped edf at the 1550*nm* regime results from spontaneous and stimulated emissions. Spontaneous radiation allows photon emission and absorption-reemission processes at different wavelengths in any direction, where only a small part of those emitted photons, $\kappa \sim 0.01 - 0.02$ remains in the fiber [23]. In a stimulated radiation the emitted photons replicate the wavelength and the direction of the impinging photons and therefore most of them are kept in the fiber. Below an <u>overall</u> inversion, the population in the second level $N_2$ is on average lower than in the first state $N_1$. However, there are wavelengths regions with $\sigma_{21}(\lambda) > \sigma_{12}(\lambda)$ where the stimulated emission increases. Therefore, the photon gas starts at the pump input in the fiber with spontaneous and stimulated emissions, and as the pump is depleted after a short distance, the photons propagate at highly non-inverted environment, undergoing spontaneous absorption-emission cycles that lead to thermalization and also stimulated emissions that replicate the spectrum. Therefore, it is possible to reach thermalization in an open fiber as the photons propagate along it, as shown in the experimental part, although most of the photons from spontaneous emission radiate out of the fiber and are lost.

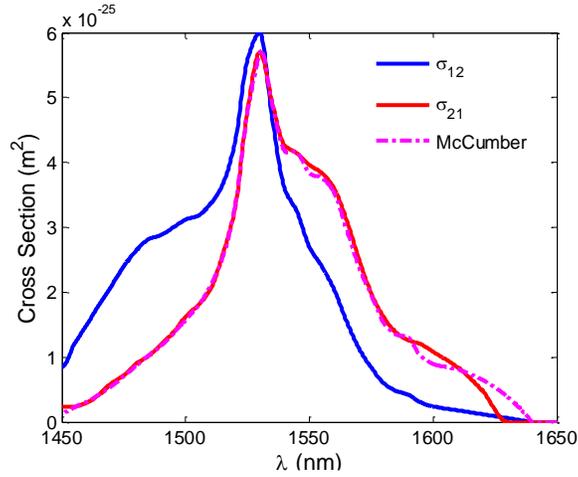

**Fig. 1:** Absorption $\sigma_{12}$ and emission $\sigma_{21}$ cross-sections of erbium in silica fibers.

A bosonic system in thermal equilibrium has a Bose-Einstein (BE) distribution, here the spectrum, $p(\varepsilon) \propto g\varepsilon / [e^{\beta(\varepsilon-\mu)} - 1]$, where $\varepsilon = h\nu = hc/(n_r\lambda)$ is energy, $h$ is the Planck constant, $\beta = k_B T$, $T$ is temperature, $k_B$ - the Boltzmann constant and $\mu$ - the chemical potential. $g$ is the light modes degeneracy or their density of states (DOS) which is independent of $\nu$ for a 1D photon gas with a regular linear dispersion. Usually the photon number is not conserved and $\mu = 0$, as we have in black-body radiation. In optical cavities and lasers, the photon number is in many cases conserved by pumping that compensates for the unavoidable losses. However, for TE and BE distribution there are additional conditions. In the dye-filled microcavity experiment [1,2], for example, the photon system, that had to meet certain strict conditions, was considered a grand-canonical ensemble with $\mu \neq 0$. In our work we find that many of those requirements are relaxed. TE is observed here in regular 1D fiber lasers, in the lasing regime, and even in open fibers without the need of micron-size cavities, high finesse, or a very high capture of the spontaneously emitted photons in all directions and a cutoff frequency [1,2].

We mentioned the possibility of lasing wit**h**out an overall inversion, but we right away stress that it only means that the total second level population is lower than the first one, and not at the specific lasing wavelength. Such lack of inversion terminology for the overall two states is used in former works [1-4,16,17]. However, one can correctly argue that for the specific lasing wavelength there is inversion. In our case, it occurs even when the pumping is between the two states at ~1550*nm* and the photon thermalization spreads the spectrum and transforms photons from low to high wavelengths where the emission cross section is larger than the absorption and compensates for the lower upper-state population. It is therefore a temperature dependent effect that results from the difference between the erbium emission and absorption cross sections, $\sigma_{12}(\lambda)$ and $\sigma_{21}(\lambda)$, shown in Fig. 1. It is possibly the case of the lack of an overall inversion in the micro-cavity BEC experiment [1,2,6-8]. We can see right away from the rate equation (Eq. 2) for $p(\nu)$ or $n(\nu)$, the possibility of T-LWI, $N_2 < N_1$, at wavelengths with a larger emission than absorption cross section, when

$$N_2 B_{21}(\nu) > N_1 B_{12}(\nu). \qquad (1)$$

We note again that $N_1$ and $N_2$ are the overall levels populations in each of the two broadened levels, and $\sigma_{21}(\lambda)$ is related to $B_{ij}(\omega)$, as done in some former works [16,17]. The lower level population at the upper state is simply compensated by the higher emission than absorption rate. Therefore, even without an overall inversion, the stimulated emission can be larger than the absorption and when it increases beyond the balance with the cavity losses it provides gain and lasing or condensation in the case of very low losses. We show below T-LWI at the high wavelength side of the thermal spectrum, ~1605*nm*, an unusual wavelength for edf fiber lasers.

## 3. EXPERIMENT AND RESULTS

We turn to the experiment and the measurements with edf ring cavities and open fibers and after that we discuss the theoretical results. We used edf with various erbium concentrations and gain figures: 30*m* of 11*dB/m*, 300*m* of 1*dB/m* and 30*m* and 100*m* of 30*dB/m*. It is important to emphasize right away that the results were quite similar for the four different fibers, and we show here only a few spectra. The pumping was at a 980$nm$ wavelength except for one case that it was with amplified spontaneous emission (ASE) of edf at the 1550$nm$ region. The ring edf cavity had regular laser losses mostly from connectors and the output coupler of 10%. It is therefore a low finesse cavity, but since it is (30-300)$m$ long it has a relatively long photon lifetime $t_c \sim (10^{-6} - 10^{-7})s$ and can be regarded as a low-loss cavity [3,4]. The mean free path $l_p$ for successive photon-erbium interaction (emission-absorption cycle) depends on the erbium concentration $N_d$ (usually of a few 100-1000 $ppm$, $N_d \sim 10^{25} m^{-3}$) and the wavelength, and is given by $l_p \approx [\sigma_{12}(\lambda) N_d]^{-1} \sim (0.1-1)m$ that corresponds to a time scale of $t_p \approx l_p /(c/n_r) \sim (0.5-5)ns$. (In silica fibers $n_r \approx 1.444$ at $\lambda \sim 1550nm$). Therefore, thermalization that needs a few photon-erbium interaction cycles [1,2] can occur within a few meters, even in open fibers, that is quite surprising. It is surely the case in our fiber cavity systems where we have $(t_c/t_p) \sim 20 - 2\times 10^3$ interaction cycles, sufficient for photon thermalization at most of the relevant wavelength region. We note that in the high-finesse dye-filled microcavity the photon lifetime was much smaller (sub $ns$) because the cavity length was only $l \approx 1.5 \mu m$, but the mean free path there was also small. The edf cavity case can therefore fall in the low-loss category in Eqs. 3 and 4, which distinguishes the BE-BEC regime from lasers [3,4,6-8], (but such low loss situation occurs in many lasers.) Nevertheless, as was already mentioned, there is another essential loss part below an <u>overall</u> inversion in our fiber systems due to the low fraction $\kappa \approx 1-2\%$ of the stimulated emission photons that is kept in the fiber cavity, while the rest radiates at all directions [23]. It means that there is a large loss factor of 98-99% at each spontaneous emission event. This loss does not stop the thermalization as we find in our experiments that are supported by the analysis and the theoretical spectra shown below. This loss factor $\kappa$ appears in the numerator of the BE distribution (Eqs. 3, 4) and simply reduces the overall intensity, but does not change the distribution. The important condition is $t_p \ll t_c$ which can hold both in long fibers and in microcavities.

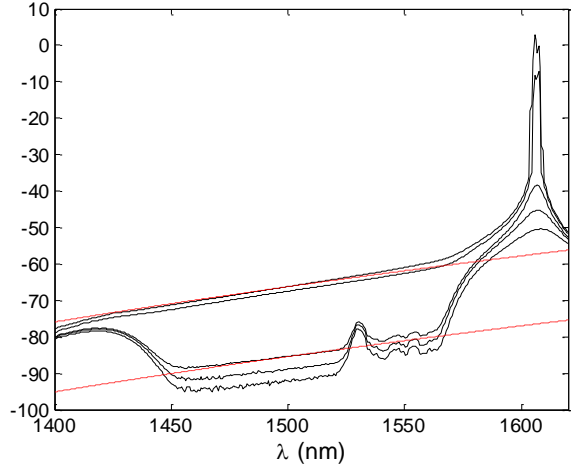

**Fig. 2:** Experimental spectra for a fiber cavity at a room temperature (~300*K*). The measured spectra (black lines) correspond to pumping of 44.67, 51.2, 53.58, 53.83, and 208.0 *mW*. The measurements shown here were taken from a ring fiber cavity consisting of 30*m* of 11*dB/m* polarization maintaining edf. The dashed red lines are the theoretical thermal BE distribution for 300*K*. The BE spectral rabge (the almost straight line parts in the semi-log plots) for low pumping are at (1450-1530)*nm* and for higher pumping at (1420-1570)*nm*. For strong pumping we see at the right side of the BE spectra a strong lasing line.

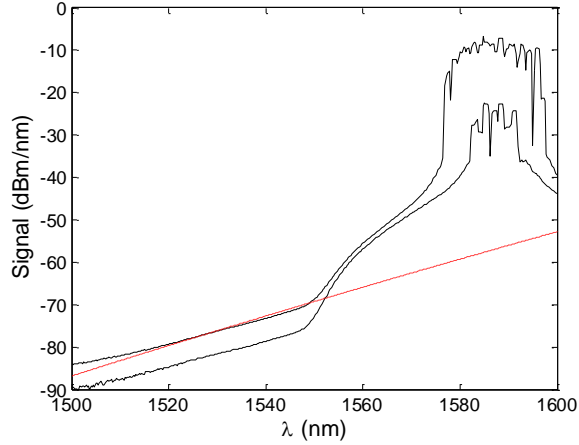

**Fig. 3:** Experimental spectra in an edf fiber cavity at 77*K*. The measured spectra correspond to pumping levels of 12 and 162 *mW*. The measurements shown here were taken from a ring fiber cavity consisting of 30*m* of 11*dB/m* polarization maintaining edf. The BE bands (the straight line parts). The dashed red line is the theoretical thermal BE distribution for 77*K*. The BE band is at (1500-1550)*nm*. For strong pumping we can see at the right side of the BE spectra strong lasing lines.

Figs. 2-3 give experimental spectra with corresponding BE line given by (Eq. 4):

$$p(\nu) \propto \kappa g\, h\nu / (e^{\beta[h\nu - \mu]} - 1) = [\kappa g\, hc / (n_r \lambda)](e^{\beta[hc/(n_r \lambda)] - \mu} - 1).$$

They show BE distribution (the almost straight line parts in the semi-log plots) at a broad spectral range up to 50-200*nm*, and with high pumping a coexistence of BE spectra with lasing without an overall inversion at the high wavelength side, around $\lambda \approx 1605 nm$. Figure 2 gives experimental spectra for an edf ring cavity measured at a room temperature, ~300*K*, and Fig. 3 at a liquid Nitrogen temperature, 77*K*. They show broad thermal spectra (the straight line parts) which nicely match the BE distribution at a broad wavelength region for both temperatures. At 300*K* for low pumping, the BE band reaches ~90*nm* and for higher pumping it spans over ~150*nm*. The lasing peaks are at the right side of the BE spectral band. At 77*K* the logarithmic BE slope is $300/77 \approx 3.9$ times higher then the 300*K* slope, as the theory predicts, but the spectral range is only ~50nm and the lasing is slightly shifted to a lower wavelength. We don't elaborate here on the effect of the chemical potential that can be disregarded in most of the band, except for the edge, as seen in the experimental spectra, but becomes zero (equals to the edge frequency) upon condensation. It is more important for the BEC study that we will discuss in a future paper on BEC. Figure 4 shows experimental spectra for an open 100*m*, 30*dB/m* edf, at a room temperature. In this experiment the BE spectral band reaches ~200*nm* and in the cavity case it is accompanied by a sharp oscillation line at the right side, that is again T-LWI. Here we used a different kind of pumping directly from the first to the second level by ASE of an edf fiber with a spectrum centered around 1550*nm*. It is a kind of "white" light pumping that doesn't generate inversion of the overall broad two levels but can give oscillation and lasing!! In the laser case this pumping was inside the cavity. It is striking to see the broad BE spectral spreading out of the narrow spectrum pumping. The thermalization process spreads and transforms the photons from low to higher wavelengths. Thus, TE can be reached not only in close cavities, but also in open fibers, as the photons propagate and become thermlized. In the distributed rate equation model that we calculated but don't show here, we could see how the photon thermalization develops along the propagation in the fiber as the distance is larger than $l_p$.

   The TE is obtained significantly below an overall inversion, but when the pumping was increased in the cavity case, lasing started without an overall inversion at an unusually long wavelength of ~1605*nm*, as seen in Fig. 2. It is the high wavelength side of the broad TE band. We observed in some cases bistability (seen in Fig. 2) and hysteresis. Our calculation showed that it can be attributed to a satuarable absorber mechanism at the fiber section where the pump is very much depleted, and due to strong backwards ASE of the edf, but it can be eliminated by a moderate pump variation along the fiber. We also note that one has to be careful about the meaning of oscillation without an <u>overall</u> inversion, as we noted above), when attributing it to lasing. We will report on this topic in the future.

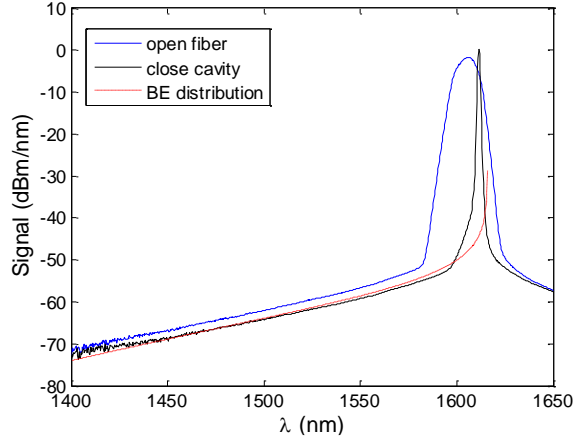

**Fig. 4:** Experimental spectra from an open fiber (blue line) and a cavity (green) at a room temperature. It was obtained with a 100*m*, 30*dB/m* edf. The BE spectral band (almost straight lines) spans over ~200*nm*, besides a sharp oscillation line (T-LWI) for the cavity case at the right side. The dashed red line shows the BE distribution at 300*K* with setting the zero frequency with the chemical potential at ~1615*nm*. The pumping here was with ASE of edf while the pumping in all other experiments was with the usual 980*nm* wavelength.

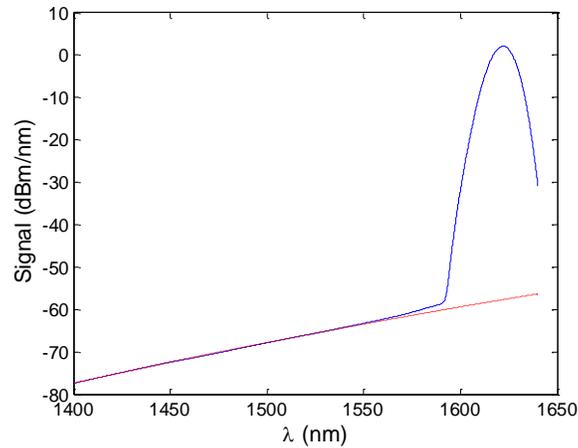

**Fig. 5:** Theoretical spectra upon propagation in an open fiber (100*m*, 30*dB/m*): at *z*=100m, calculated by the rate equation.

## 4. THEORETICAL ANALYSIS

The theoretical analysis is based on the lumped or the distributed rate equations models, given in the Methods part. We show in Fig. 5 the theoretical spectrum of the thermalization effect on the light that propagates along an (100*m*, 30*dB/m*) open fiber. We can see the similarity to the experimental result in Fig. 4. Figure 6a shows broad BE spectra (>120*nm*) calculated by the lumped model for a low cavity loss (high photon lifetime $t_c$). Similar spectra can be obtained by the distributed model. The results are in a good agreement with the experiment, showing a large band (>120*nm*) of thermal spectrum. These figures also show T-LWI with typical values of $N_2/N_d \sim 0.18$, that is much below an <u>overall</u> inversion. The theoretical figures show a very good match to the experimental spectra.

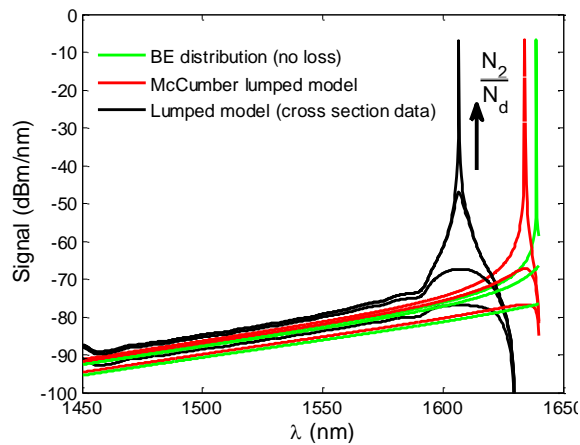

**Fig. 6a:** Theoretical spectra of a fiber cavity at 300*K* with a broad BE regime (the straight lines in the semi-log plots), calculated by the lumped rate equation model (Eq. 2) for a low-loss cavity with a photon life-time $t_c = 10^{-6} s$. The black lines correspond to solutions of the rate equation for increasing values of the population ratio until the threshold value: $N_2/N_d = 0.09, 0.162, 0.179, 0.18$. The figure also shows spectra calculated from the rate equation with a cross section enforced by the McCumber relation. The red lines are for $t_c = 10^{-6} s$ and the green lines for $t_c \to \infty$ (no cavity loss), with the above inversion and a slightly different threshold value), showing oscillation at the band edge.

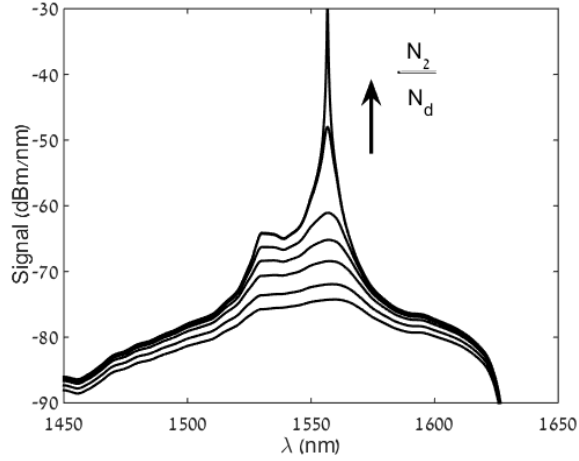

**Fig. 6b:** Theoretical spectra for a regular edf laser oscillation. Calculated by the lumped rate equation model (Eq. 2) for 300$K$. It is the relatively high cavity loss regime or short photon lifetime, here with $t_c = 2.3 \times 10^{-9} s$. The spectrum here is very different from the low loss thermal BE regime and the lasing is at 1556.6$nm$. The lines correspond to increasing values of the population ratio until their threshold value: $N_2/N_d = 0.31, 0.372, 0.434, 0.496, 0.558, 0.6193, 0.62$.

The drastic difference between the thermal lasers and regular lasers spectra can be seen by comparing Figs. 6a and 6b. As mentioned above, the thermal lasers with broad BE spectra have low cavity losses. We used in Fig. 6a typical photon lifetime of $t_c = 10^{-6} s$ that fits our fiber lasers, and also show the spectrum for zero loss ($t_c \rightarrow \infty$). This broad BE spectra is very different from Fig. 6b that shows the spectra of a regular laser that has a higher cavity loss with $t_c = 2.3 \times 10^{-9} s$. It is in accordance with the theoretical study that differentiates TE and BEC from lasers by losses [3,4,7,8], except that the spontaneous emission loss is very high in our systems. Remarkably, the very simple lumped model fits our experiments, the dye microcavity [1,2,7-8], regular lasers and photon-BEC behavior.

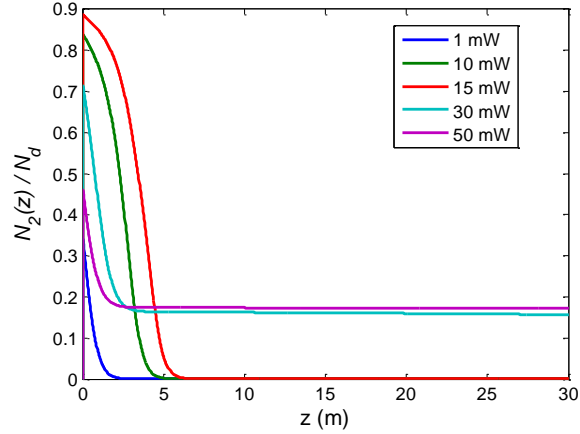

**Fig. 7:** Theoretical upper state population fraction $N_2(z)/N_d$ along an edf fiber for various pumping levels. Calculated by the distributed rate equations model (Eq. 5). Below lasing (blue, green and red curves) the $N_2/N_d$ drops down to near zero values after a short propagation in the fiber. Above the lasing threshold (purple and sky blue curves) the upper state population fraction is almost constant along the fiber and approaches an average value of 0.18 that is much below inversion, showing the T-LWI phenomena.

Fig. 7 shows the overall upper level population ratio $N_2/N_d$ as the photons propagate in the fiber, obtained from the distributed model (Eq. 5). This model includes an equation for the pump that implies its depletion after a short distance in the edf. It shows that the upper state population quickly falls to a small value. In the case with the oscillation, the population ratio stabilizes at a finite value well below an overall inversion (~0.18 for the experiment parameters). We note that besides the input and output measurements of the pump and the signal, and therefore the inversion, their variations along the fiber are obtained by the theoretical analysis, as done in former works, such as in the microcavity experiment, [1,2,6-8].

## 4. THE RATE EQUATIONS MODEL

We elaborate here on the theoretical base of the lumped and distributed rate equations. The lumped model gives the basic results shown in the theoretical figures that nicely describe the experimental spectra. The second distributed rate equations model allows following the pump and signal propagation along the fiber (*z*-dependence).

## A. The lumped rate equation

We first describe the lumped model that gives the basic results. As in the dye microcavity system [7-8], the photon population and its dependence on frequency is governed by a rate equation where the erbium atoms are modeled as broadened two level systems:

$$dn(\nu)/dt = \kappa N_2 A(\nu) + N_2 B_{21}(\nu)n(\nu) \\ - N_1 B_{12}(\nu)n(\nu) - n(\nu)/t_c. \quad (2)$$

The Einstein A-B relation here takes the form $A(\nu) = B_{21}(\nu)g(\nu)$, with $g(\nu)$ as the light modes degeneracy or DOS (constant in 1D with a linear dispersion). The steady state solution of the rate equation is

$$n(\nu) = \frac{\kappa g(\nu)}{\frac{t_c^{-1} + N_1 B_{12}(\nu)}{N_2 B_{21}(\nu)} - 1}. \quad (3)$$

For $t_c^{-1} \ll N_1 B_{12}(\nu)$ with the McCumber [18-20] (or Kennard-Stepanov [21-22]) relation $B_{12}(\nu)/B_{21}(\nu) = \sigma_{12}(\nu)/\sigma_{21}(\nu) = Ce^{\beta h\nu}$ where C is constant, Eq. 2 reduces to the BE distribution and spectrum:

$$n(\nu) = \kappa g(\nu)/(e^{\beta(h\nu-\mu)} - 1) \\ p(\nu) \propto n(\nu)h\nu = \kappa g(\nu)h\nu/(e^{\beta(h\nu-\mu)} - 1). \quad (4)$$

The chemical potential is defied by $e^{\beta\mu} = CN_1/N_2$, $\sigma_{ij}(\nu) \propto B_{ij}(\nu)$ are the absorption and emission cross sections $\sigma_{12}(\lambda)$ and $\sigma_{21}(\lambda)$ related to transitions between the first and second levels (shown in Fig. 1), and $B_{ij}(\omega)$ are the Einstein coefficients.

## B. The distributed rate equation

The theoretical base for T-LWI in our system is derived from an extension of the above model to distributed (z-dependence) rate equations for the populations, where the erbium atoms are modeled as broadened two level systems. It gives more insight on the dynamics inside the cavity and includes information about the propagation of the signal and the pump along the fiber z-axis. The model consists of the following set of coupled equations (in the two-level approximation) for the pump and signal powers, $P(z)$ and $p(\nu, z) \propto n(\nu, z)h\nu$:

$$dP(z)/dz = -\sigma_p N_1(z)$$
$$dp(\nu,z)/dz = [\sigma_{21}(\nu)N_2(z) - \sigma_{12}(\nu)N_1(z)]p(\nu,z) \quad (5)$$
$$+ \kappa N_2(z)\sigma_{21}(\nu)g(\nu)h\nu.$$

$$dN_2(z)/dt = \sigma_p N_1(z)\frac{P(z)}{h\nu_p A_p} + \int d\nu \sigma_{12}(\nu)N_1(z)\frac{p(\nu,z)}{h\nu A_s}$$
$$- \int d\nu \sigma_{21}(\nu)N_2(z)\frac{p(\nu,z)}{h\nu A_s} - \frac{N_2}{t_{sp}} = 0,$$

where $n(\nu,z)$, $N_1(z)$ and $N_2(z)$, $N_1(z)+N_2(z)=N_d$ are photon, population and erbium atom densities, respectively.

The first equation describes the pump $P(z)$ absorption by the ground level atoms with a cross section $\sigma_p \approx 2.6 \cdot 10^{-25} m^2$. The starting condition is $P(z=0)=P_0$. The second equation describes the propagation of the signal spectrum $p(\nu,z)$, governed by stimulated absorption and emission, and on the loss $\kappa$ of the spontaneous emission. The starting condition for this equation depends on the fiber geometry: $p(\nu,0)=p_{in}(\nu)$ for open fibers, and $p(\nu,0)=R\,p(\nu,L)$ for close cavities, where $R$ is the output coupler reflection. The third equation describes the population dynamics of $N_{1,2}(z)$. $A_p$, $A_s$ are the pump/mode overlap areas with the fiber core ($A_p, A_s \approx 24\mu m^2$), and $t_{sp} \approx 10^{-2} s$ is the spontaneous emission time.

## 5. CONCLUSION

We conclude by a note on the possibility for experimental BEC observation in edf cavities, which would be a most important and ambitious goal for a photon gas in 1D at a room temperature. A major step for it is thermalization of the photon gas that is obtained in the present work. There are a few other soluble issues, such as a requirement on the density of light-mode states in 1D cavities, particle (photon) number conservation that can be achieved by pumping, while being in the low-loss quantum (non-laser) regime that we believe is achieved in the present work, reaching a critical power for BEC that we estimate to be in the $\mu W - mW$ region, and the need of a cutoff frequency that determines the ground-state for the condensation. The linear dispersion in 1D fiber cavities near and above the frequency cutoff gives a frequency independent density of light-mode states (DOS) that is at the boundary of BEC (logarithmic) transition, that should be enough for a finite system. Small deviation to sublinear dispersion that gives DOS with a nonzero positive power needed for a full BEC transition is present in various doped fibers, including our edf, and can be also obtained by other dispersion engineering methods, such as the use of nonlinearly chirped gratings.

Another important point learned from the present work is that a lack of an overall inversion in the broad edf two level transitions doesn't eliminate the possibility of lasing. This can be the case in the dye-filled microcavity experiment [1], and therefore lasing there, and not necessarily BEC, should not be excluded. Nevertheless, there were other points that can add clearance on the BEC observation [1,6-8], and also questions about it [9]. We don't elaborate here on this point but stress that BEC and its relation to lasing in such broad level systems need additional study and discussion that we shall provide in a future publication.

**Funding**. This research was supported by the Pazy Foundation.

**REFERENCES**

1. J. Klaers, J. Schmitt, F. Vewinger. and M. Weitz, "Bose-Einstein condensation of photons in an optical microcavity," Nature **468**, 545-548 (2010).
2. J. Klaers, F. Vewinger, and M. Weitz, "Thermalization of a two-dimensional photonic gas in a white wall photon box," Nat. Phys.**6**, 512-515 (2010).
3. P. Kirton, aqnd J. Keeling, "Nonequilibrium model of photon condensatation," Phys. Rev. Lett. **111**, 100404 (2013).
4. P. Kirton, and J. Keeling, "Thermalization and breakdown of thermalization in photon condensates," Phys. Rev. A **91**, 0332826 (2015).
5. R. Weill, A. Bekker, B. Levit, M. Zhurahov, and B. Fischer, "Breaking two laser "axioms": Lasing without inversion and thermal equilibrium," arXiv 1607 01681.  ( https://arxiv.org/pdf/1607.01681 ).
6. J. Klaers, J. Schmitt, T. Damm, F. Vewinger, and M. Weitz, "Bose-Einstein condensation of paraxial light," Appl. Phys. B, **105**, 17-33 (2011).
7. J. Schmitt, T. Damm, V. Dung, F. Vewinger, J. Klaers, and M. Weitz, "Thermalization kinetics of light: From laser dynamics to equilibrium condensation of photons," Phys. Rev. A **92**, 011602 (2015).
8. J. Klaers, Schmitt, J. Schmitt, T. Damm, F. Vewinger, and M. Weitz, "Statistical Physics of Bose-Einstein-Condensed Light in a Dye Microcavity," Phys. Rev. Lett. 108, 160403 (2012).

9. B. Fischer, and R. Weill, "When does single-mode lasing become a condensation phenomenon?" Opt. Express **20**, 26704-13 (2012).
10. R. Weill, B. Levit, A. Bekker, O. Gat, and B. Fischer, "Laser light condensate: Experimental demonstration of light-mode condensation in actively mode locked laser," Opt. Express **18**, 16520-25 (2010).
11. R. Weill, B. Fischer, and O. Gat, "Light-mode condensation in actively mode-locked lasers," Phys. Rev. Lett. **104**, 173901 (2010).
12. G. Oren, A. Bekker, and B. Fischer, "Classical condensation of light pulses in a loss trap in a laser cavity," Optica **1**, 145-148 (2014).
13. M. Zhurahov, A. Bekker, B. Levit, R. Weill, and B. Fischer, "CW laser light condensation," Opt. Express **24**, 6553 (2016).
14. C. Sun, S. Jia, C. Barsi, S. Rica, A. Picozzi, and J. W. Fleischer, "Observation of the kinetic condensation of classical waves," Nat. Phys. **8**, 470-474 (2012).
15. C. Connaughton, C. Josserand, A. Picozzi, Y. Pomeau, and S. Rica, "Condensation of classical nonlinear waves," Phys. Rev. Lett. **95**, 263901 (2005).
16. C. Conti, M. Leonetti, A. Fratalocchi, L. Angelani, and G. Ruocco, "Condensation in disordered lasers: Theory, 3D+1 simulations, and experiments," Phys. Rev. Lett. **101**, 143901, (2008).
17. A. Fratalocchi, "Mode-locked lasers: Light condensation," Nat. Photon. **4**, 502-503 (2010).
18. E. Desurvire, "Erbium-doped fiber amplifiers," Wiley Publication (1994).
19. D. E. McCumber, "Theory of phonon terminated optical masers," Phys. Rev. **134**, 299 (1964).
20. R. M. Martin, and R. S. Quimby, "Experimental evidence of the validity of the McCumber theory relating emission and absorption for rare-earth glasses," J. Opt. Soc. Am. B, **23**, 1770 (2006).
21. E. H. Kennard, "On the Interaction of Radiation with Matter and on Fluorescent Exciting Power," Phys. Rev. **28**, 672 (1926).
22. B. I. Stepanov, "A universal relation between the absorption and luminescence spectra of complex molecules," Dokl. Akad. Nauk SSR **112**, 839 (1957).
23. T. Sondergaard, and B. Tromborg, "General theory for spontaneous emission in active dielectric microstructure: Examples of fiber amplifier," Phys. Rev. A, **64**, 033812 (2001).